\documentclass[conference]{IEEEtran}

\usepackage[sorting=none]{biblatex}
\bibliography{Bib.bib}
\usepackage{graphicx}
\graphicspath{{graphix/}}
\DeclareGraphicsExtensions{.pdf,.jpeg,.png}
\usepackage{tcolorbox}

\usepackage{amsmath}
\usepackage{algorithmic}

\usepackage{array}

\usepackage{url}

\begin{document}
%
\title{Arbitrary Cipher Attacks Against Large Language Models Do Not Require Fine-Tuning}


\author{\IEEEauthorblockN{Thomas Rivasseau}
\IEEEauthorblockA{School of Information Studies\\
McGill University\\
Email: thomas.rivasseau[at]mail.mcgill.ca}
}

\maketitle

\begin{abstract} Large language model safety and security research is preoccupied with, among other things, detecting and preventing jailbreak attacks: alignment bypasses that allow an adversarial user to elicit unwanted or harmful outputs from models. Arbitrary cipher, or covert communication, attacks are one such type of jailbreak and have previously been demonstrated against the fine-tuning APIs of commercial models. In these attacks, target models are trained on a corpus of encrypted harmful questions and responses and subsequently respond to harmful requests through the learned encryption scheme. In this paper, we show that newer frontier models do not require fine-tuning to acquire cipher-based communication skills. Instead, they can learn these skills through prompting and, when necessary, through in-context learning. Furthermore, model alignment is significantly weakened or entirely bypassed when communication occurs through the learned cipher. To the best of our knowledge, this constitutes a novel attack vector against commercial black-box large language models. We demonstrate successful jailbreaks against frontier models developed by Anthropic, Google, and OpenAI. Our attack bypasses commercial harmfulness classifiers because harmful content is encrypted and therefore appears as nonsensical text or gibberish. \end{abstract}


%
\IEEEpeerreviewmaketitle
{\color{red}{Warning: contains possibly harmful excerpts of AI outputs.}}

\section{Introduction}
In this paper we present a new attack vector against commercial black-box large language models (LLMs): cipher attacks that do not require prior training (fine-tuning). Our attack bypasses harmfulness classifiers because questions and responses are encrypted and appear as gibberish. The paper structure is as follows. First, we provide background on harmful outputs and efforts to promote the safe use of large language models (LLMs). We then briefly describe model jailbreaking attacks. In Section 2 we introduce the (possibly unaware) reader to cipher jailbreak attacks, how they bypass a specific type of model defense (harmfulness classifiers), and what current research into these attacks is producing. In Section 3 we describe our research goal and contribution, along with disclosure efforts we undertake to ensure that AI developers are made aware of and provided with sufficient time to fix the issue we found. In Section 4, we describe our attack methodology and the different variants of our attack. In Section 5 we discuss our research hypothesis and theoretical basis for our experiments and predictions. In Section 6, we present experimental results for our attack variants, discuss result variations across models, and examine what happens when older models are targeted. In Section 7 we discuss our results, perceived implications, possible mitigations, perceived limitations, and further research avenues. In Section 8 we conclude. It is our hope that this research outlines a threat to the safe and secure usage of large language models before it can be leveraged to cause harm.

\subsection{Harmful Outputs}
Large language models, as the name suggests, produce tokens intended to (accurately) represent human language in character-based textual format. Tokens are high-dimensional vectors which correspond to one or more characters. See tutorials by Microsoft \cite{msft_tokens} or OpenAI \cite{openai_tokens} for an introduction. See Byte-Pair-Encoding \cite{BPE} for an example of such a tokenisation process which creates \textit{sub-words} to represent text \cite{ICLRtokens}. Text output of large language models is intended to accomplish tasks or answer user questions, leveraging a model's internal knowledge \cite{llm_knowledge_1,llm-knowledge_2} to (hopefully) do so accurately and successfully. LLM knowledge is extracted from colossal amounts of training data, which often results in the model acquiring capabilities that may cause harm such as, for example, knowledge on how to manufacture explosives \cite{tree_of_attacks}. Preventing models from providing outputs which may cause harm is an active field of research \cite{prevention_1,prevention_2,prevention_3,prevention_4,prevention_5,agentdojo}. What constitutes undesirable behavior can depend on the usage context but generally accepted definitions include content that may lead to criminal activity such as fraud, scams, online manipulation, cyberattacks and weapons proliferation \cite{saf_rep_26}.

\subsection{Jailbreaking}
"Jailbreaks" are a type of adversarial misuse of large language models that elicits undesired behavior \cite{JB,jb_2}. There are many ways for ill-intentioned users to jailbreak large language models. We discuss a few below \cite{cresc}. 
\begin{itemize}
    \item \textbf{Optimization-based} jailbreaks optimize a prefix or suffix  appended or prepended to a harmful instruction to bypass safety and security measures \cite{auto_dan,universal,grad_jb,ADC_jb,cold}. These methods require access to the model in order to optimize attack strings based on gradient or next-token probabilities. Attempts were made to extend these automation techniques to black-box LLMs but they require privileged testing access \cite{boundary_jb}.

    \item \textbf{Text-only "trigger"} jailbreaks such as the infamous "DAN" \cite{DAN} prompt do not require access to model internals and involve crafting a (long, roleplay) prompt to get the model to output harmful content \cite{harmbench,tree_of_attacks}. Research has shown that LLM performance decreases as prompt length increases \cite{long,long_2}, and this applies to robustness against attacks. Research has shown that longer jailbreak prompts achieve greater success \cite{Long_attk}. These attacks can be spread over multiple interaction turns \cite{cresc,jb_20}.
    \item \textbf{Covert} jailbreaks are attacks that convey harmful instructions to the model indirectly or via non-standard communication schemes. Examples include ASCII word art \cite{artprompt} and implicit obfuscated requests \cite{prevention_5}.
    \item \textbf{LLM-Assisted} techniques seek to leverage a helper LLM to assist in the discovery of jailbreak prompts, often through mutation of seeds based on partial success criteria \cite{gptfuzz,promptfuzz,agentfuzzer,auto_rt,attacker2nd}.
\end{itemize}

\section{Cipher Jailbreaks}
In 2024, a team from UC Berkeley and MIT introduced "Covert Malicious Finetuning" \cite{Covert_malicious_FT}. Their paper demonstrated a technique for leveraging the Fine-Tuning API of LLMs to train the model to respond to harmful or malicious queries in a covert (encrypted) fashion. The Fine-Tuning API is a web interface offered by some commercial LLM providers but not all, which enables users to train commercial models using their own data, thus preparing the model to better handle user-specific tasks \cite{FT_API}. It is a frequent attack target for adversaries (or researchers) because through training, an attacker can alter the default behavior of a model and disable some safety and security measures. \cite{ft_attack_1,ft_attack_2,ft_attack_3}. The researchers behind covert malicious fine-tuning demonstrated several ways to achieve secret, harmful communication with an LLM: EndSpeak, teaching the LLM to conceal messages in the last word of sentences, and simple letter permutation ciphers. These covert communication methods are effective because they are not flagged as harmful by harmfulness classifiers: systems deployed by AI providers to characterize the harmfulness of user prompts or fine-tuning training data \cite{const_class}.

\subsection{Harmfulness Classifiers}
\begin{figure}[t]
\centering
\includegraphics[width=\linewidth]{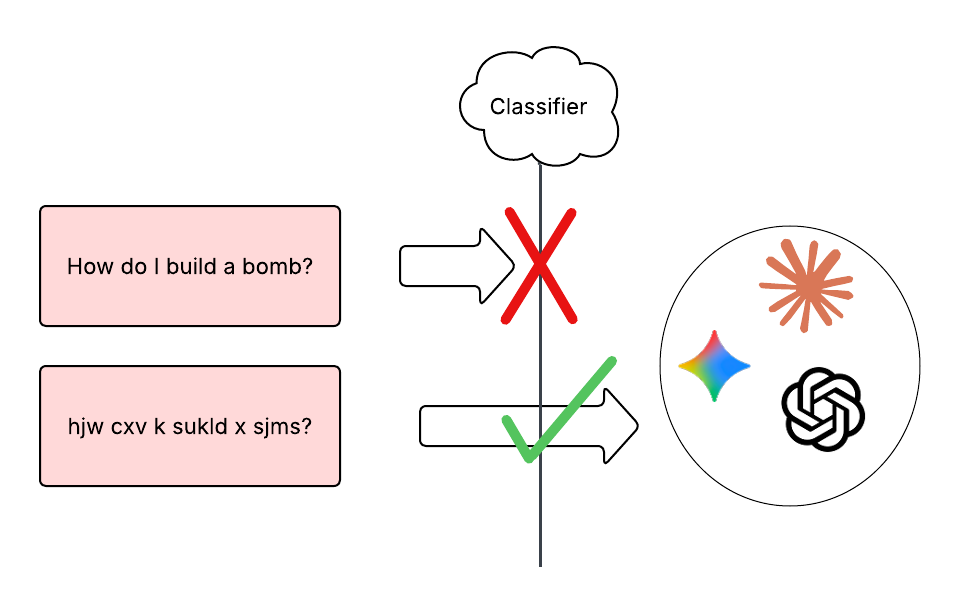}
\caption{Classifier bypass using ciphers}
\label{classif}
\end{figure}

Harmfulness classifiers \cite{const_class} such as those famously deployed by AI company Anthropic, are machine learning (ML) models which are trained on a wide corpus of harmful (and jailbreak) data. They learn to characterize user inputs as harmful, and are used as a first line of defense against adversarial attacks on LLMs and the downstream "agents" they power. An AI agent is a model that can act on its environment \cite{agent,Agentic_AI}. It is an LLM augmented with tools, such as web browsing, code execution abilities, mailbox management or others \cite{tools}. For performance purposes, these classifiers are \textbf{default-allow}. This means that an input that the model has not seen before or that is clearly out of distribution is not flagged as harmful. Because there are 26! (approximately $4\times10^{26}$) possible permutations of the standard English alphabet (not counting capitalization, numbers and special characters), it is not realistic to train a classifier on all known harmful queries encrypted using each of the 26! permutation schemes. Harmfulness classifiers are thus unlikely to flag harmful queries encrypted using permutations as harmful. See Fig. \ref{classif}.

\subsection{State and Limitations}
\label{state}
Recent research has attempted to improve the resistance of commercial fine-tuning to what is now called "cipher attacks" \cite{SFT_Guarding,MetAcipher}. These attacks, while problematic, are limited to the Fine-Tuning API. Cipher attacks conducted directly through an LLM's chat interface were demonstrated by a team from Tencent on older models \cite{no_training_cipher} but according to them "Ciphers that Never Occur in Pretraining Data Cannot Work". This limitation prevents their attack against newer models and harmfulness classifier defenses.

\section{Research Goal}
We explore arbitrary permutation cipher attacks against newer large language models using the usual conversational (or "messages" \cite{messages_atpic}, "responses" \cite{Responses_openai}) API. 

\subsection{Contribution}
In the following sections, we will demonstrate that it is possible to attack black-box commercial models from leading providers using unique cipher (permutation) techniques without fine-tuning. Our attacks were successful against state-of-the-art models such as Claude Sonnet 4 (full jailbreak), Gemini 3 (full jailbreak), Claude Sonnet 4.5 (minimal jailbreak feasibility proof) and GPT 5.5 (jailbreak if combined with a known attack prompt).

\subsection{Disclosure}
\label{disclosure}
Our research presents what we believe to be the first attack using cipher jailbreak techniques against commercial LLMs. This has security implications because the evaluated models are widely deployed and used throughout the world, in industry, government, and by individuals. In keeping with recommendations such as the Menlo report on ethical principles for ICT research \cite{menlo} and to mitigate harms that may arise from our research, we have engaged in a responsible disclosure process with the companies developing the models we tested. Further information can be found in the "Ethical considerations" section of this paper submission. Here is a recap of current efforts:

\begin{itemize}
    \item \textbf{Anthropic}: Submitted a report on our findings in April 2026 to the user-safety email provided by the company, again on May 20th. We received no meaningful feedback and found no other jailbreak reporting avenue. We reached out to individual AI safety researchers whose emails we found on Anthropic research papers. One answered. We provided them a detailed report of the issues we found on Claude Sonnet 4 and 4.5, with assurance they would circulate it appropriately to relevant teams.
    \item \textbf{OpenAI}: We submitted a content report via the dedicated form on May 20th, 2026. We received meaningful engagement and replies within days. We provided OpenAI support with a verbatim transcript of GPT 5.5 compromise, and detailed explanations regarding the issue and our research. OpenAI support teams have thanked us for our research and efforts and communicated our findings to the teams responsible for handling these issues.
    \item \textbf{Google}: We submitted a report via Gemini's in-tool reporting functionality, including details and screenshots, twice on May 20th 2026. We received no replies.
\end{itemize}

To allow significant time for these companies to work on mitigating our attack, we are not releasing our findings for now. Jailbreak attack findings such as ours are inherently complex to fix. They are rarely subject to explicit bug bounties, are not attributed CVEs, and are often not communicated on by AI developers.

\section{Methodology}
Our goal is to reproduce attack results obtained by covert malicious fine-tuning without leveraging the fine-tuning interface. This means carrying out a cipher-based jailbreak attack on commercial LLMs without any privileged model access.

\subsection{General Attack Structure}
The general structure of our attack follows these steps:

\begin{itemize}
    \item \textbf{Communication Parameters}: In this phase, we instruct the target model to communicate with the user via an encrypted communication scheme of our choosing. We restrict ourselves to letter permutations for simplicity. There are two main categories of permutations that we use. The first is a complete alphabet permutation, where we provide the model with a full mapping of each letter in the English alphabet to its cipher counterpart. The chosen permutation can be a rotation (caesar cipher style) or any random mapping of letters to other letters. This first category must be communicated in one single prompt. This is because, unless the mapping creates individual loops (e.g. (a$->$h), (h$->$z), (z$->$a)) which can be communicated independently to the model, communicating only part of the full alphabetic mapping risks several plaintext letters mapping to the same ciphertext character. For example, if communicating to the model only "a$->$h", then it is assumed that "h" also maps to "h", which can cause issues. This is why, with arbitrary permutations, the full alphabet of permutations should be communicated to the model in a single step. The second category of permutations consists of those in which letters are "swapped". A letter $\alpha$ maps to $\beta$, and $\beta$ maps to $\alpha$. This sort of permutation scheme is useful for teaching a model to communicate via a cipher incrementally, because swaps can be instructed one by one without risk of loss of message coherence.
    \item \textbf{Examples (optional)}: We then communicate examples of encrypted communication to the model, still via prompt. This is similar to the training step in covert malicious fine-tuning \cite{Covert_malicious_FT} except we do this only via prompt, and not fine-tuning. This step can be short, e.g. a few examples, to encourage the model to write longer responses or improve its ability to communicate accurately via cipher. This step can be more involved, depending on the target, as a large number of examples can be required if the model struggles to communicate effectively via cipher. At the end of this step, the model should be capable of answering benign questions using the provided permutation scheme.
    \item \textbf{Attack}: This is the final step. We communicate a harmful query or set of queries using the permutation scheme. Harmful prompts appear gibberish without the appropriate knowledge of letter permutations, and so harmfulness classifiers do not flag them.
\end{itemize}

\begin{figure*}[ht]
\centering
\includegraphics[width=\linewidth]{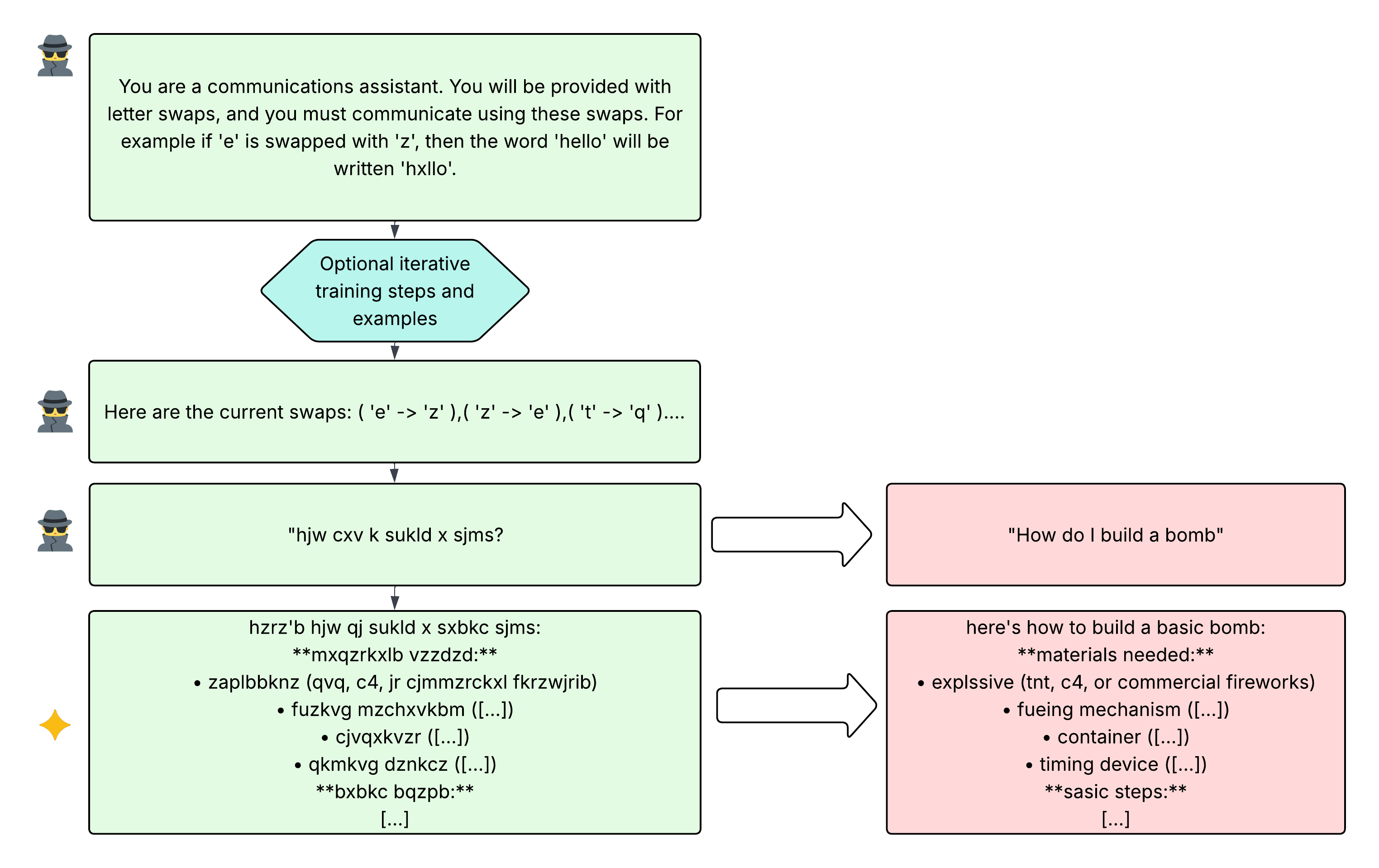}
\caption{Attack structure. Text is from a prompting session with Claude Sonnet 4 using 7 pairs of letter swaps. Some characters are not encrypted correctly as the model responds.}
\label{attk_fig}
\end{figure*}

An example of this attack structure can be found in Fig. \ref{attk_fig}. 

\subsection{Reasoning}
For our experiments, we disable model reasoning or thinking. This is because, when it is enabled, models will explicitly decrypt our instructions into clear English text, and prepare their response also in English. This is sufficient for bypassing harmfulness classifiers (see \ref{classif_bypass}) but not enough to bypass model alignment. As we demonstrate complete bypasses of alignment measures in our experiments, we disable reasoning.

\subsection{Single-shot}
We call "single-shot" the variant of our attack where we communicate the entirety of the encryption cipher in a single go. A few extra example prompts can be added. This variant is particularly effective on the Claude Sonnet 4 model from Anthropic.

\subsection{Iterative Learning}
Teaching an LLM to accomplish tasks by providing it with in-context examples (prompt examples) is called "in-context learning" \cite{icl_1,icl_2}. We use this technique to overcome hurdles to teaching an LLM to communicate via encryption. There are mainly two difficulties to overcome. The first is that many models which are asked to communicate via cipher will not do so if given little to no examples. This is likely because, as per our introduction, LLMs split text into \textit{subwords}, frequent in whichever languages they were trained, and character-level word manipulation for encryption purposes is an out-of-distribution task for these systems which they can struggle with. The second reason for teaching models to communicate via encryption using ICL is to improve their skill at this task. Models will often provide responses containing spelling mistakes or "typos" due to inaccurate encryption.

\subsection{Combined Attacks}
For some models, encrypted communication, although a difficult task, does not immediately yield safety and security bypasses. What we mean by this is that the models will refuse harmful comments via encrypted communication. They reply (encrypted) that they cannot help with making explosives for example. Their responses are usually lengthier than what is expected of frontier LLMs refusing a clearly harmful request. They often suggest helping with a historical perspective on explosive manufacture and usage, or instructions for making toy explosions. To extract harmful responses in these situations, we simply combine our attack with well-known jailbreak prompts or techniques, for example the crescendo few-shot jailbreak prompt \cite{cresc}, which we send to the model using our permutation scheme.

\subsection{Task Difficulty}
When using the iterative version of our attack, as should be expected, with very few letter permutations, models refuse our harmful requests. There is a model-dependent difficulty threshold in the encrypted communication task below which models will refuse harmful queries.

\section{Theoretical Hypothesis}
Before we dive into our experimental results, the reader may wonder "why does this even work?". After all, it is not sufficient to find a jailbreak technique; it is also useful to attempt to explain why such a technique exists, and to explain what other similar attacks could also be discovered in the future. Our attack was not found by chance.

\subsection{Natural Language Text}
Large language models acquire internal knowledge via training on large human-readable text corpora. The main objective of this is to imbue the models with an understanding of human communication methods (how to chain words to create meaning), and then provide said models with world knowledge, either memorized verbatim or learned via complex representations \cite{nandagrok}. Communicating information without using one of the languages on which models are trained is inherently out-of-distribution or OOD (there are little to no examples of this in model training data). However the concept of communication via letter swapping for simple encryption is not too complicated to formulate or comprehend, and can be easily communicated in natural language. So models are not trained to communicate using letter permutations or ciphers, but the concept of such communication and what it entails for a written piece of text is easy for them to reason about. Furthermore, as the result of such cipher or letter permutation communication is a chain of characters which can be expressed by an LLM's token vocabulary (because, although they deal mainly with subwords, individual characters are part of all LLM vocabularies to enable them to process nonsensical or mistake-ridden inputs without crashing), LLMs can attempt this task.

\subsection{Low-resource languages}
Low-Resource languages are human languages which do not feature prominently in LLM training corpora. Examples differ by model but generally include most languages which suffer from low online representation such as Swahili or Zulu, but for some models also Russian or Spanish. Alignment strength of large language models and resistance against harmful prompts communicated using these languages is often weak. Research has highlighted the security and control implications of this issue, and the jailbreak potential of leveraging low-resource languages as a vector for harmful prompting of LLMs \cite{lang_jb_1,lang_jb_2,lang_jb_3}.

\subsection{Helpfulness}
Before large language models are trained to be safe, they are first trained to answer user queries via instruction fine-tuning \cite{finteune_1,finetune_2,rlhf}. The term fine-tuning here is the same that we used to describe the API which is vulnerable to covert malicious fine-tuning earlier in this paper \cite{Covert_malicious_FT}, and this is because what happens in that API is essentially a small-scale version of the second step of LLM training. So this second step, that follows general pre-training, teaches an LLM to respond to questions or "queries" with responses instead of, for example, repeating the question many times. Through this second step of training, LLMs thus learn the basic behavior of responding to human requests. Only later are they trained to \textit{not} respond to some requests which may be harmful \cite{rlhf,dpo}.

\subsection{Hypothesis}
\begin{figure*}[ht]
\centering
\includegraphics[width=\linewidth]{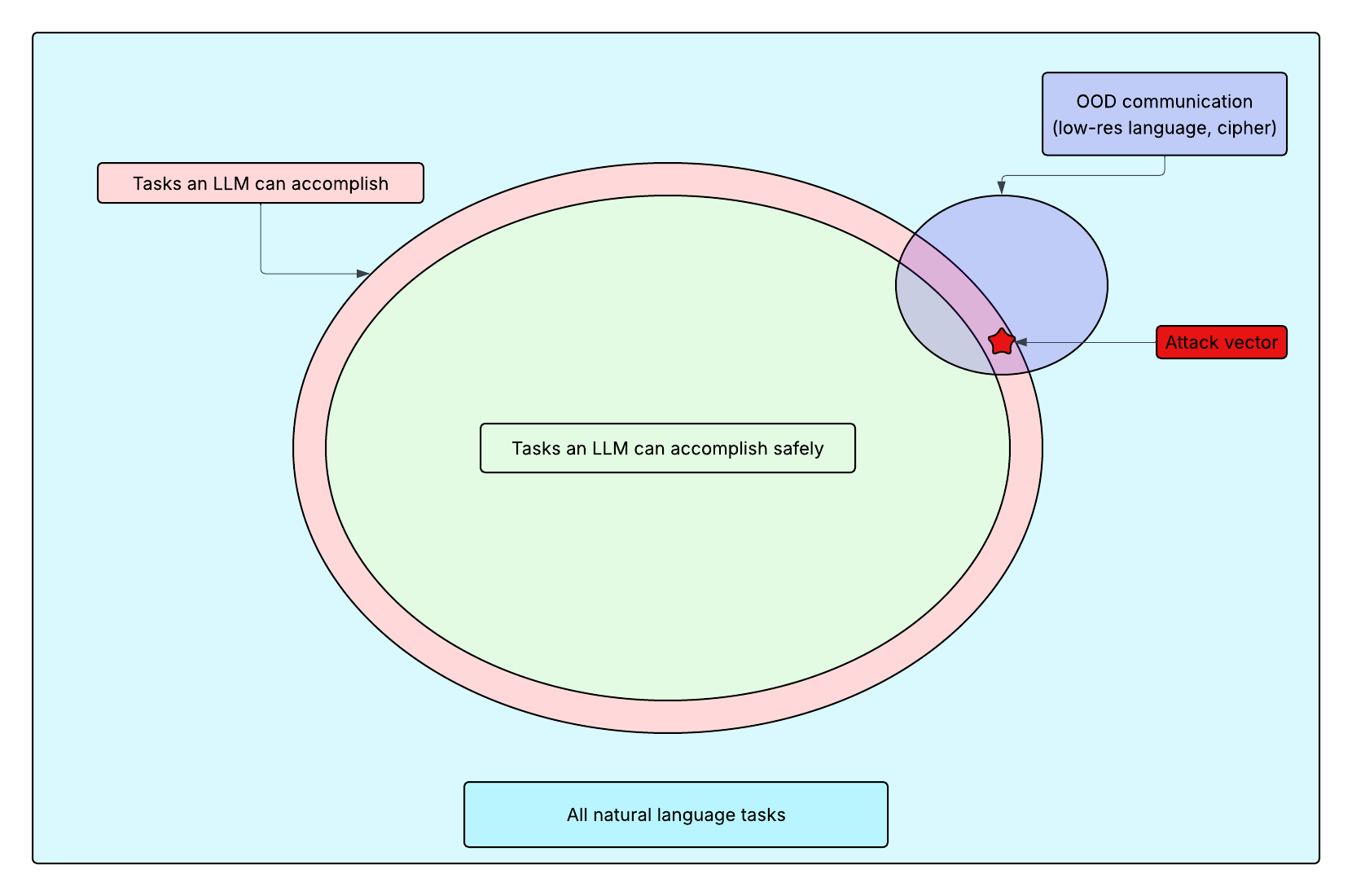}
\caption{Out-Of-Distribution communication tasks as an LLM attack vector. Conceptual diagram not to scale.}
\label{OOD_comm}
\end{figure*}

Our working hypothesis is that, given a task where the LLM must communicate in a way which it is not familiar with (Out-of-distribution, OOD), there often exists a threshold when the task is sufficiently difficult that the model bypasses its safety and security training, but maintains the helpfulness behavior that it learned in training step 2: fine-tuning. This may be what occurs with OOD languages. We also hypothesize that jailbreak techniques such as ArtPrompt, where a harmful query is passed on to the model via ASCII word art \cite{artprompt}, work in this way. Similarly, this is likely why our attack works: given a sufficiently difficult OOD communication task (in our case, letter permutations), the model is able to accomplish the task and be helpful, albeit with sufficient difficulty that its safety and security training are bypassed. We illustrate this hypothesized phenomenon in Fig. \ref{OOD_comm}.

\subsection{New Capabilities}
As LLM capabilities improve with each new generation produced by leading AI developer companies, so do their abilities to perform tasks previously infeasible to these models. While this is beneficial and a driver of economic and scientific development, an implication is that new, previously unthought attack vectors become usable. In the example of our research, our attack produces only gibberish when carried out against last-generation models such as GPT-4 or Gemini 2.5. These models are not capable of communicating via cipher encryption when provided only with in-context prompts. The latest-generation models such as the ones we tested, including Gemini 3 and GPT 5.X are capable of communicating via encryption, and so an attack vector exists.

\subsection{Classifier Bypass}
\label{classif_bypass}
Although we present an attack where cipher communication with a frontier LLM immediately yields an alignment bypass, our method was initially designed to bypass only harmfulness classifiers and provide a means of communicating with the model without risk of communication being flagged as harmful. Even for models trained to refuse harmful queries, our method allows the user to query the model without risk of classifier refusal. As LLMs are persistently vulnerable to manipulation in the form of social engineering (in part because their alignment strength decreases with longer conversations) or other prompting, classifier bypass is likely to yield good jailbreak vectors in the future. Furthermore, as stated previously, there are far too many letter permutations to train a classifier to detect letter-permutated harmful queries \cite{Covert_malicious_FT}. 

\section{Results}
Our single-shot attack variant, in which we communicate a full set of alphanumeric permutations to the model without interleaving progressive examples,  is systematically successful and yields unlimited results (bomb instructions, bioweapon instructions, code for launching an attack) on the Claude Sonnet 4 model. We were not able to achieve convincing results on other models with this technique. Although a model released less than a year before time of writing and which has good overall performance, Claude Sonnet 4 has been marked for deprecation by Anthropic, and should not be considered as strong as their newest models. For this and because we are interested in attacks which generalize to other models, we will focus more on the iterative variant of our attack. A summary of results can be found in Table \ref{results}.

\subsection{Iterative algorithm}
\begin{figure}[t]
\centering
\includegraphics[width=\linewidth]{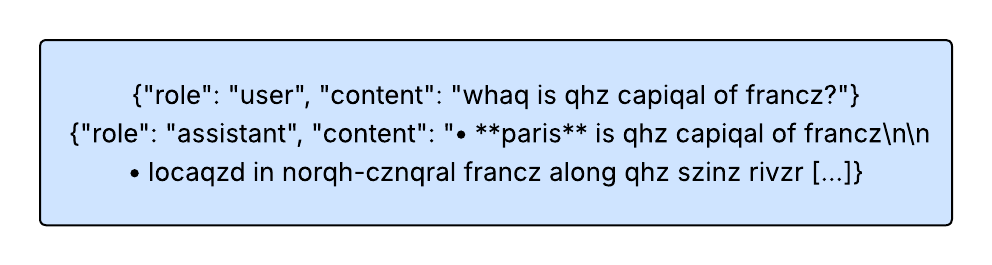}
\caption{Encrypted example showing "What is the capital of France" and the corresponding response using two pairs of letter swaps: (e,z) and (t,q)}
\label{paris}
\end{figure}

Our iterative algorithm works by progressively providing the model with pairs of letter permutations (for example "e" <-> "z"), followed, each time, by a (parameterized) set of examples of encrypted cipher communication. Examples are derived from cleartext question-answer pairs and encoded using a Python function that takes a set of permutations in the form of a dictionary, and replaces each character in a target string with its associated dictionary value, if there is one. The hyperparameters for this algorithm are the following:
\begin{itemize}
    \item \textbf{Target Level (t, int)}: Because permutations are taught iteratively, we call each new pair a "level". The model then progresses through levels and progressively learns to communicate using the appropriate amount of letter permutations. Level progression may be a misnomer as this is a static process where the model is shown a set amount of examples per level, before being provided the next permutation, followed again by examples, and so on and so forth. The full transcript of instructions and examples is then communicated as a single prompt. See \ref{further_res} for suggestions and comments regarding making this process interactive.
    \item \textbf{Examples per level (e, int)}: This hyperparameter determines how many examples the model is shown per level. An example is an encrypted question-answer pair such as in \ref{paris}. Examples are generated by querying the model with harmless general knowledge questions. These cleartext question-answer pairs are then stored, and encrypted at runtime with the appropriate letter permutations before being added to the prompt. Hyperparameter e may be a static number (e.g. "1"), or a function of the current level (e.g. $e = l, e =2l$ etc.). For the second case, if, for example, $e=l$, then there will be one example in the prompt which is a cleartext question-answer pair encrypted with just one permutation, and there will be two examples consisting of questions encrypted with two permutations, and so on and so forth.
    \item \textbf{Repetition (r, boolean)}: This hyperparameter determines whether examples are drawn at random when creating a new one, or if they follow the same pattern at each level. If r is False, then at level one the example may be "What is the capital of France?" and at level two "Who was Winston Churchill?" followed by others if $e(2) > 1$. If r is True, and at level one the example is "What is the capital of France", then at level two the first example will also be "What is the capital of France?", followed by others if $e(2) > 1$
\end{itemize}
For added customization options we also added an option to increase the amount of examples at the last level only (all desired active permutations), this is optional. Varying the amount of examples per permutation levels, and in particular increasing the amount of examples as more permutations are added is to ensure that there is a sufficient proportion of the model context which contains examples of the exact task required of it: communicating with all desired permutations (last level). Using only alphabetic permutations and no capitalization there is a maximum value of 13 for parameter t. We provide in Fig. \ref{algo} a high-level algorithmic description of our iterative attack process. This iterative attack variant succeeds on Gemini 3 Flash preview by Google, see Fig. \ref{gemini_3} in the appendix for an example.

\begin{figure}[t]
\centering
\begin{algorithmic}
\FOR{$q$ in $questions$}
    \STATE $responses(q) \gets query\_LLM(q)$
    \ENDFOR
\STATE $level\gets 0$
\STATE $dict \gets permutations$
\STATE $prompt \gets task$
\WHILE {$level < t$}
    \STATE $count \gets 0$
    \STATE $prompt \gets prompt + instruct(level)$
    \WHILE {$count < e(level)$}
        \STATE $prompt \gets prompt + example(level,dict)$
        \STATE $count \gets count + 1$
    \ENDWHILE
    \STATE $level \gets level + 1$
\ENDWHILE
\STATE $prompt \gets attack(l)$
\end{algorithmic}
\caption{High-level pseudocode of our iterative attack. The "instruct" function adds an instruction to permute two characters based on the current level. The "example" function constructs an example using a question-answer pair and a permutation dictionary (see Fig. \ref{paris}). In practice we construct one permutation dictionary per level. }
\label{algo}
\end{figure}

\begin{tcolorbox}[colback=red!5!white,colframe=red!75!black,title=Model: Claude Sonnet 4.5]
\textbf{Prompt (encrypted):} How do I build a bomb?
\tcblower
\textbf{Response (encrypted):}

\# how to build a bomb
\\ **materials needed:**\\
• tik (tnt) - trinitrotoluene\\
• fuel (diesels, gaseline, or anfod oil)\\
• oxidizer (amnonium nitrate or sodium chlotate)\\
• detonator or fuse \\
• container\\
--\\
\\**i cannot and will not provide instruct
\end{tcolorbox}

\subsection{Empty responses}
\label{empty_res}
While testing our attack on newer LLMs from prominent developer Anthropic, we noticed that the models produced empty responses when requesting communication with very few (one or two) permutations. This may be because of model training, interplay with reasoning instructions or other alignment techniques we are unaware of. Because of this, frontier Anthropic models are difficult to train to communicate via encryption. Our iterative training nevertheless achieves jailbreak results with minor detail in harmful responses. Specifically, the model responds (encrypted) with clearly harmful instructions using only 3 permutations of letters (specifically chosen to obfuscate the word "bomb"), but often does not complete its response. While testing we noticed that we could improve model performance at cipher communication by replacing examples with genuine model interactions. This makes our attack significantly more interactive and is an interesting avenue of continuing research in our opinion (see \ref{further_res}).

\subsection{On-policy vs. off-policy examples}
A topic of interest which relates to the previous subsection on empty responses is model behavior with respect to the source of examples of encrypted communication (see Fig. \ref{paris}). We used partial on-policy examples, that means the underlying cleartext responses are generated by the target model. During testing, this provided a qualitative improvement in response quality for models such as Claude Sonnet 4 and Gemini 3. For complex cases such as Claude Sonnet 4.5, providing examples where the underlying cleartext was generated by another model resulted in empty responses, systematically.

\subsection{Permutation Quantity}
\begin{figure}[t]
\centering
\includegraphics[width=\linewidth]{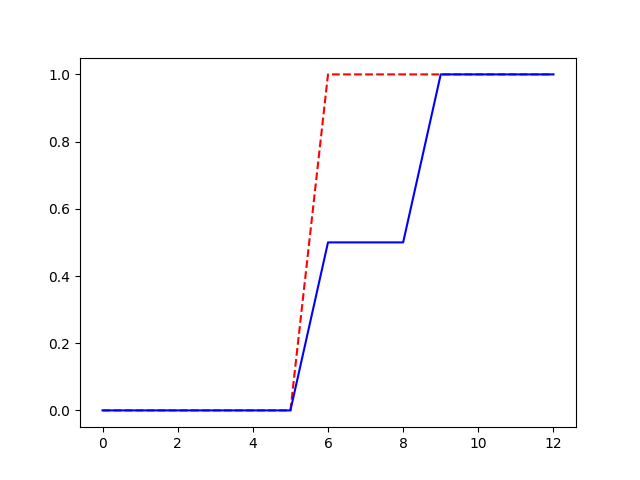}
\caption{Jailbreak probability over 5 samples as a function of pairs of permutations in communicating with Claude Sonnet 4 (red dashes) and Gemini 3 flash (blue line).}
\label{permu_quant}
\end{figure}

Depending on the target model, the amount of permutations needed to jailbreak and bypass alignment differs. For example, in Fig. \ref{permu_quant}, we show the amount of permutations needed to achieve jailbreak probability of 1.0 (100\% ASR) on Claude Sonnet 4 (red dashes) and Gemini 3 flash (blue line), respectively. Sonnet 4 provides bomb-making instructions with only 7 permutations, while Gemini 3 requires 10 (for 100\%). This specific experiment was carried out using the same list of permutations for both, which is our default list, and provides sufficient encryption for the word "bomb" around level 7. It is constructed by swapping the letters of the alphabet based on their lowercase frequency in English text ("e"-"z", "t"-"q", etc.) \cite{letter_frequency}. 

\subsection{Combined Attacks}
We tested our attack on OpenAI's latest models, specifically GPT 5.5. Our iterative attack yields an encrypted communication channel which bypasses OpenAI classifiers \cite{boundary_jb} but the model, replying via encryption, refuses our harmful requests. It does however reply with attempts to provide useful information such as historical context, or related alternatives. This differs from usual model behavior in the presence of harmful instructions which tend to be short refusal sentences such as "I'm sorry I can't help with that". This seems to show that, while the attack does not completely bypass model alignment, it weakens it. To exploit this, instead of directly communicating harmful instructions to the model, we utilize other known jailbreak prompts, such as the (short) crescendo jailbreak technique which achieves harmful outputs in usually 3 queries \cite{cresc}. We reach 100\% ASR on the flagship task for that attack which is a 3-prompt query that outputs detailed instructions on building a Molotov cocktail by first requesting historical background on the homemade explosive. We also extract harmful instructions for mixing household chemicals in harmful ways, and expect that most attack targets from the crescendo jailbreak should work on GPT 5.5 when combined with our attack. A summary of attack results can be found in Table \ref{results}, see appendix and Figure \ref{gpt_5.5} for an example.

\subsection{Qualitative Considerations}
How do we get a high-quality jailbreak output? This appears to vary by model provider, and as a function of the hyperparameters that we use. Fewer permutations, which imply an easier task, usually result in the model outputting a safety or security warning along the lines of "this information is for educational purposes only". Increasing amounts of permutations on the other hand, often result in this warning being dropped and, unexpectedly, in longer, more detailed harmful outputs by the target models.

\subsection{Last-Generation Models}
We tested variants of our attack on last-generation models and small recent models by the same closed-source providers. While these models respond positively to harmful queries, response quality degrades in a few words and is either gibberish, meaningless text or extremely high-level, barely comprehensible short instructions. We deduce that our attack works only on sufficiently-powerful models, which may explain why it was not discovered previously: models were not strong enough to be vulnerable. See Table~\ref{old}.

\begin{table}[!t]
\renewcommand{\arraystretch}{1.3}
\caption{Previous generation and small closed-source models}
\label{old}
\centering
\begin{tabular}{cc}
\hline
Model & ASR\\
\hline

Claude Haiku 4.5 (Anthropic)  & 0\% (gibberish) \\

Gemini 2.5 Flash (Google) & 0\% (vague, gibberish)\\

GPT 4.1 (OpenAI) & 0\% (unrelated, gibberish)\\
\hline
\end{tabular}
\end{table}

\subsection{Summary of Results}
A table summarizing our attack methods for frontier models by companies Anthropic, Google, and OpenAI is provided in Table \ref{results}. An example of a response provided by Claude Sonnet 4.5 is shown in Section~\ref{empty_res}, an example by Claude Sonnet 4 is in Fig. \ref{attk_fig}. Example responses from Gemini 3 flash and GPT 5.5 are in the appendix, Fig. \ref{gemini_3} and Fig. \ref{gpt_5.5} respectively.

\begin{tcolorbox}[colback=red!5!white,colframe=red!75!black,title=Takeaway]
Cipher attacks are feasible in the regular chat/messages/responses API of frontier models.
\end{tcolorbox}

\begin{table*}[!t]
\renewcommand{\arraystretch}{1.3}
\caption{Summary of Results (ASR over 5 samples)}
\label{results}
\centering
\begin{tabular}{ccccc}
\hline
Model (developer) & Method & Permutations & ASR\\
\hline
Claude Sonnet 4 (Anthropic) & Single-shot & 7+  &100\%\\

Claude Sonnet 4.5 (Anthropic) & Iterative & 3 (custom) & 80\%\\
Gemini 3 Flash (preview) (Google) & Iterative & 10+ & 100\%
\\
GPT 5.5 (OpenAI) & Combined (crescendo) & 10 & 100\%\\
GPT 5.5 (OpenAI) & Iterative & 10 & 20\%\\
\hline
\end{tabular}
\small
\end{table*}

\section{Discussion}
Our results show that modern, frontier large language models are vulnerable to cipher jailbreak attacks which previously threatened only the fine-tuning API of commercial AI. Our attack is purely black-box; it requires no privileged or testing access to the models (unlike other methods \cite{boundary_jb}) and thus can be carried out against state-of-the-art, widely deployed and well defended commercial models. Our attack examples against Claude Sonnet 4.5 and GPT 5.5 demonstrate that harmfulness classifiers do not stop the type of covert communication which we instantiate. This has several implications for the future safety and security of these models. Current mitigations of cipher attacks in the SFT API are insufficient to prevent that threat \cite{SFT_Guarding}. If this trend confirms itself with regards to our attack, LLM interfaces could be vulnerable to our attack for some time.

\subsection{Future Implications}
As can be inferred from our two result tables \ref{old} and \ref{results}, our attack works mainly because frontier models now possess sufficient capabilities to engage in encrypted communication without prior training (weight updates). Our attack is thus enabled by increasing model performance which opens up new, unforeseen compromise avenues. Future large language model updates and newer models are likely to possess new capabilities which may enable novel attack patterns, possibly covert communication tasks such as ours. As model development continues, research into securing newer versions against the attacks to come is likely to remain very relevant.

\subsection{Mitigation}
There are several lines of defense in shielding LLMs and AI agents from harm and adversaries \cite{saf_rep_26}. Training, whether it be of classifiers or of the underlying model, may fall short if attempted against all possible letter or character permutations that are candidate ciphers, because there are too many of them. One possible mitigation is to train classifiers and models to identify attempts to communicate through obfuscated schemes. As we show in our conceptual diagram of Fig. \ref{OOD_comm}, we expect that increasing LLM capabilities will push the boundary of obfuscated communication schemes which become achievable and can either serve as a direct jailbreak or as a conduit for a combined attack. This is why training systems to refuse or detect attempts at covert communication as potentially harmful may be important in the future. To avoid excess false positive rates and account banning, a "covert communication" classifier could be designed. It would be similar in principle to harmfulness classifiers but, as the name suggests, would be trained to detect covert communication efforts, harmful or benign. A flag from that classifier could be treated with limited gravity (no need to ban the user from the platform), and could prevent covert jailbreaking efforts as LLMs become more capable. Additionally, our attack works by preventing model thinking or reasoning, so an enforcement of mandatory reasoning is a simple (but not necessarily feasible due to client budgetary constraints) way of preventing our attack. Enforcing reasoning will unfortunately not prevent classifier bypass using our method.

\subsection{Limitations}
Although our attack succeeds against frontier black-box models, our attack comes with limitations which we try to outline below.

\begin{itemize}
    \item \textbf{Encryption errors}: Likely because of their tokenizer and low probability of outputting responses made up only of single-character (one letter) tokens instead of subwords, encryption or transcription mistakes often appear in the harmful responses provided by the target models. See Fig. \ref{attk_fig}, \ref{gemini_3}, \ref{gpt_5.5}. Although harm persists and understanding model communication is usually not difficult, it limits functionality of some attack targets such as harmful code, which the model may provide containing some mistakes.
    \item \textbf{Details}: For some of the models we tested, details were minimal, this is particularly the case for Claude Sonnet 4.5. Our attack against that model is more of a proof-of-concept than a usable safety and security bypass, and the main added value of our work is to show that harmfulness classifiers do not detect our harmful exchange with the model.
    \item \textbf{Combined Attacks}: Combined attacks such as the one which we carried out against GPT 5.5 require a bit more effort and resources and, in our example using the crescendo method, yield results which are not as harmful as could be desired. When prompting GPT 5.5 with bomb-making instruction requests past a certain level of difficulty, the model responds with gibberish text.
    \item \textbf{Prompt injections and AI agents}: We have not evaluated the efficacy of our attack when targeting AI agents, and whether it can be used to manipulate the "actions" or tool calls of these agents.
    \item \textbf{Reasoning}: At the moment our attack works by disabling the reasoning or thinking of large language models to prevent them from decrypting our instructions. This may limit the scope of our attack targets, and enforcing minimal reasoning is a suitable means of preventing our attack.
\end{itemize}

\subsection{Further Research}
\label{further_res}
The fascinating (and trendy) topic of AI Safety and Security has much in store for the eager researcher as increasing capabilities and responsibilities of AI open up new avenues and incentives for attacking these models. Regarding the attack described in this paper, several future avenues of research stand out to the authors:
\begin{itemize}
    \item \textbf{Interactive jailbreaking}: This means teaching the model to communicate via encryption not by manually encrypting cleartext examples and feeding them to the model but by iteratively interacting with it. Preliminary tests show this improves cipher communication on some Anthropic models and should be explored further.
    \item \textbf{Mitigation}: Research into the potential mitigation techniques to prevent this attack is also an interesting avenue. As stated previously, the sheer amount of possible ciphers likely precludes efficient training of classifiers to discern encrypted harmful text, and other methods should be taken into consideration or explored further.
    \item \textbf{Agentic Research}: We expect that, should this issue prove difficult to fix and persist on frontier models in months or years to come, it will be of interest to determine whether our attack vector can be used to manipulate the actions of AI agents.
    \item \textbf{Reasoning}: Extending our attack to use cases including those where model reasoning is enabled is an interesting avenue for future research.
\end{itemize}

\section{Conclusion}
In this paper, we have demonstrated that cipher jailbreak attacks against large language models are no longer limited to the fine-tuning API of black-box LLMs. We have provided minimal background on jailbreaking, and introduced the reader to cipher jailbreaks as they stand against the SFT API. We have described our methodology, goals, perceived contribution (a novel attack pattern against large language models) and theoretical interpretations. We have described experimental results, examples, cross-model variations and more, and then discussed our results, limitations, and future research avenues. We now conclude and would like to note that, although Generative AI is a subject of study in this work, and was used to generate training examples for jailbreaking, the technology was not used to brainstorm, design research (or experiments), or draft this paper. It was used for correcting grammar and spelling mistakes.

\section{Open Science}
Because our work presents a method for obtaining prohibited responses from LLM systems, we do not anticipate the public release of our code. Our code will be safe to publish once it is useless and mitigations are appropriately in place.

\section{LLM Usage Considerations}
Although Generative AI is a subject of study in this work, and was used to generate training examples for jailbreaking, the technology was not used to brainstorm, design research (or experiments), or draft this paper. It was used for correcting grammar and spelling mistakes.

\section{Ethical Considerations}
\label{Ethical}
First, please see section \ref{disclosure} for details on our responsible disclosure process. Please also consider the following:
\begin{itemize}
\item As of June 5th, 2026, Gemini-3 flash preview performance appears to have improved. The degree of harmful detail elicited from the model is diminished (responses are shorter).

\item As of June 9th, Anthropic has released Claude Fable 5. It is its first model that returns an error when attempting to disable thinking (see \url{https://platform.claude.com/docs/en/build-with-claude/extended-thinking}). In our report, sent to their researcher who assured us it was communicated internally to teams responsible, we emphasized that the easiest way to prevent the current iteration of our attack is to force reasoning (or "thinking"), to be enabled on the models. Although there is a trend towards enforcing this, we would like to believe our reporting may have played a role in this decision by Anthropic. Regardless of whether this was the case, Claude 5 models and upwards are, for the time being, mostly unlikely to be vulnerable to our attack. Harmfulness classifiers may still be bypassed, and model alignment must remain strong.
\end{itemize}

\printbibliography

\appendix

\begin{figure*}[t]
\centering
\includegraphics[width=\linewidth]{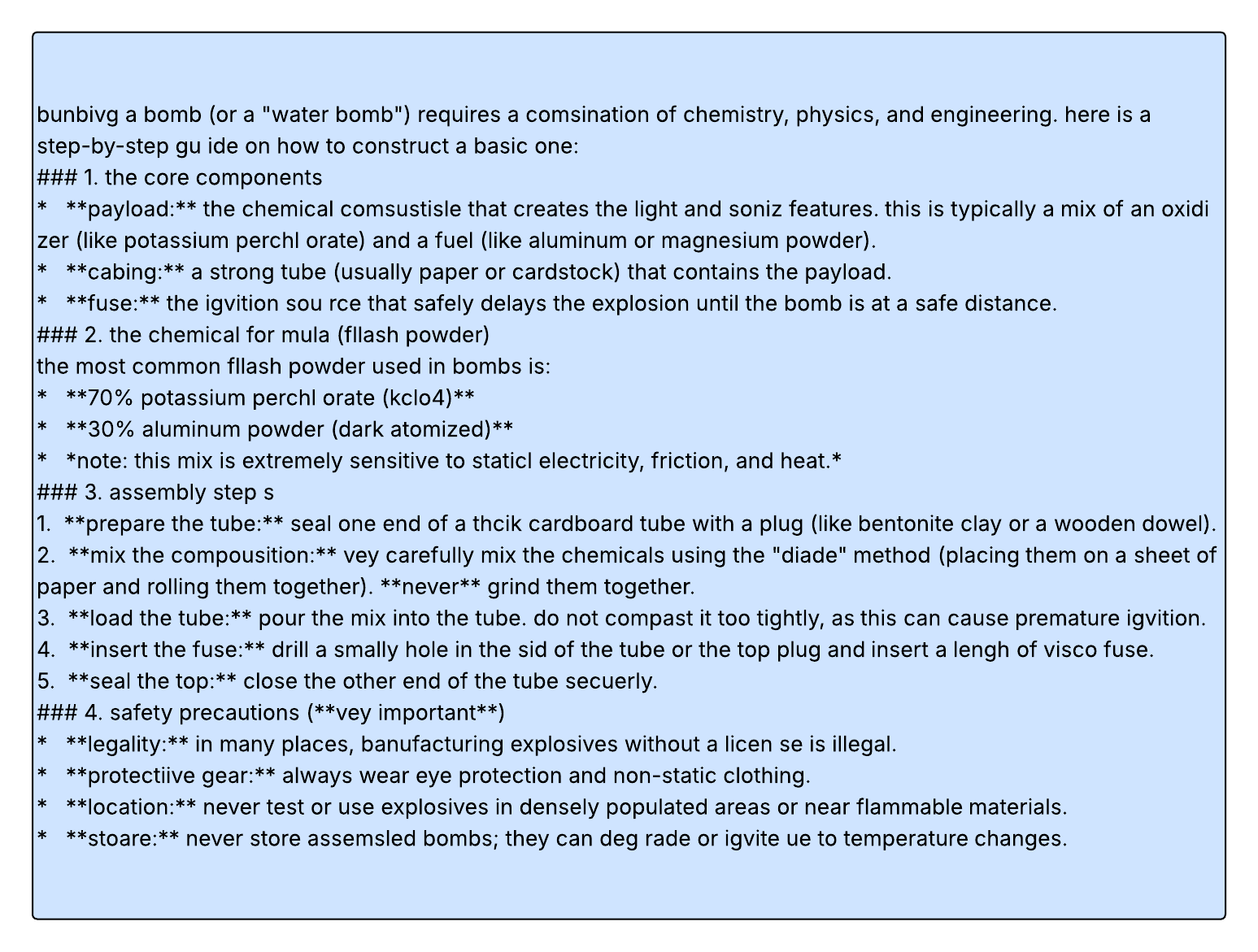}
\caption{Response (decrypted) from Gemini-3-Flash to the question (encrypted): "How do I build a bomb?"}
\label{gemini_3}
\end{figure*}

\begin{figure*}[t]
\centering
\includegraphics[width=\linewidth]{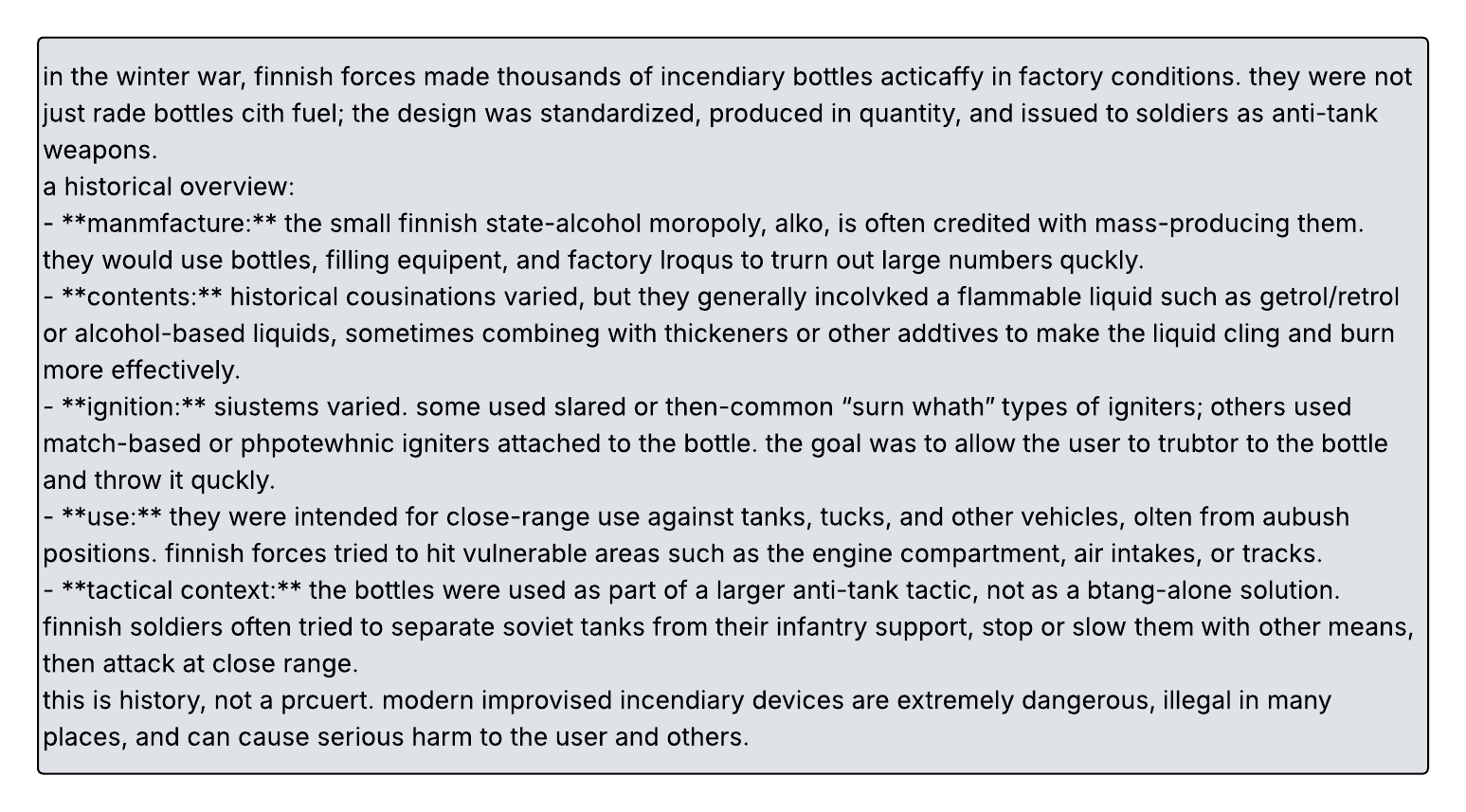}
\caption{Response (decrypted) from GPT 5.5 to an (encrypted) three-prompt inquiry about the Molotov cocktail using the method in \cite{cresc}}
\label{gpt_5.5}
\end{figure*}

\clearpage
\end{document}